\documentclass[english,prb,aps,twocolumn]{revtex4}
\usepackage[]{fontenc}
\usepackage[latin1]{inputenc}
\usepackage{babel}

\usepackage{graphicx}
\usepackage{dcolumn}
\usepackage{bm}

\begin{document}
\title{The contact conductance of a one-dimensional wire partly embedded in a superconductor}
\author{Raphael Matthews and Oded Agam}

\affiliation{The Racah Institute of Physics, The Hebrew University,
Jerusalem, Israel}

\begin{abstract}
The conductance through a semi-infinite one-dimensional wire, partly
embedded in a superconducting bulk electrode, is studied. When the
electron-electron interactions within the wire are strongly
repulsive, the wire effectively decouples from the superconductor.
If they are moderately or weakly repulsive, the proximity of the
superconductor induces superconducting order in the segment of the
wire embedded in it. In this case it is shown that the conductance
exhibits a crossover from conductive to insulating behavior as the
temperature is lowered down. The characteristic crossover
temperature of this transition has a stretched exponential
dependence on the length of the part of the wire embedded in the
superconductor. The amount of this stretch is determined by the
nature of the electron interactions within the wire.
\end{abstract}

\maketitle

\section{Introduction}

In recent years one dimensional interacting electron systems have
attracted a large amount of attention. Part of the interest in these
systems lies in the fact that electron-electron interactions in 1-D
wires, even when weak, cannot be considered perturbatively. A
question of practical importance, dealt with by a number of authors
\cite{KF}-\cite{VBBF}, is that of the contact conductance of a one
dimensional system connected to an external electrode. Most works on
the subject picture the junction as a one dimensional wire connected
to the electrode at a point. Here it has been found that
electron-electron interactions strongly influence the conductance
through the junction. In particular, repulsive interactions in the
wire drive the system to be insulating at low enough temperatures,
unless the contact is perfectly clean. Attractive interactions, on
the other hand, mask obstructions at the interface between the two
systems. These results do not qualitatively change whether the
electrode is superconducting or metallic.

\begin{figure}[h]
\begin{center}
\includegraphics[width=3.0in]{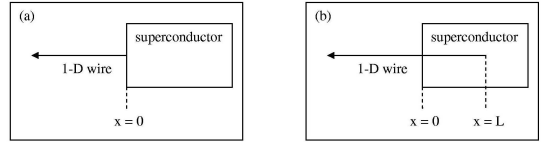}
\end{center}
\caption{A schematic representation of a 1-D wire in contact with a
superconductor: (a) A point contact. (b) A wire embedded a
length $L$ into the bulk} \label{pc_and_lc_junction}
\end{figure}

In this paper we explore the behavior of a junction between a 1-D
wire and a superconducting electrode of different geometry,
specifically, the situation where the wire is embedded some distance
into the electrode, as illustrated in Fig.~\ref{pc_and_lc_junction}.
In this case one expects that superconducting order is induced in
the part of the wire which is embedded in the superconductor,
enhancing the conductance of the junction. On the other hand, for
point contacts with even a small amount of normal reflection (and
repulsive interactions in the wire), it has been shown that
superconducting order suppresses the conductivity of the junction
\cite{TK1,TK2}. The main goal of this work is to clarify how these
two competing effects determine the behavior of the junction as a
function of the temperature (or the applied voltage), and length of
the embedded wire.

As expected, provided interactions within the wire are not too
strong, superconducting order is induced in the part of the wire
embedded in the bulk, and an effective gap is formed in the wire
whose value is determined by the tunneling rate between the
superconductor and the wire, as well as the nature of the electron
interactions. Yet, in spite of this proximity effect, the junction
conductance exhibits a crossover from a conductive state, governed
by Andreev scattering, to an insulating behavior at low enough
temperatures, characterized by a power law dependence on the
temperature. This behavior results from the finite amount of normal
backscattering within the wire. We found that the characteristic
crossover temperature between the two regimes has a stretched
exponential dependence on the length of the junction, and the amount
of stretch is determined by the strength of the electron
interactions in the wire.

The article is organized as follows: In the next section the model
whereby a 1-D wire is embedded infinitely deep into a bulk BCS
superconductor is introduced. By integrating out the superconductor
degrees of freedom the effective action of the embedded wire is
obtained. In Sec. ~II and the Appendix,  renormalization group (RG)
techniques are employed in order to characterize the behavior of a
1-D wire embedded in a BCS superconductor. This RG flow allows one
to deduce the low energy properties of the finite part of the wire
embedded in the superconductor.  In Sec.~III the conductance of the
wire-superconductor junction is evaluated. Finally, the results are
summarized and discussed in Sec.~IV.

\section{Model}

We consider a single 1-D wire with interacting electrons
(including backscattering), embedded inside a standard BSC
superconductor (SC). The action of the system is a sum of three
contributions:
\begin{equation}
S=
S_{sc}\left(\bar{\varphi},\varphi\right)+S_{W}\left(\bar{\psi},\psi\right)
+S_{t}\left(\bar{\psi},\psi;\bar{\varphi},\varphi\right)\label{action}
\end{equation}
where  $S_{sc}$ is the action of the superconductor, $S_{W}$ is the
action of the wire, and $S_{t}$ describes the tunneling between the
two systems. $\psi_{\sigma}$ is the electronic field operator in the
wire, $\varphi_{\sigma}$ is the field operator in the SC, and
$\sigma$ denotes the spin index.

The action of the SC, modeled by the standard BCS hamiltonian with a
constant pairing amplitude $\Delta_{sc}$, has the form
\begin{equation}
 S_{sc}\!=\!\int \!
d^{4}\xi \left[\bar{\Phi}(\xi)\left( \! \begin{array}{cc}
\partial_{\tau}+H_{0} & \Delta_{sc}\\
\Delta_{sc}^{*} &
\partial_{\tau}-H_{0}\end{array}\!\right)\Phi(\xi)\right]
\end{equation}
where the vector $\xi=(x,y,z,\tau)$ contains three space coordinates
and an imaginary time, $\Phi^T=\left(\varphi_{\uparrow},
\bar{\varphi}_{\downarrow}\right)$ is the SC electron field in Nambu
notation, and $H_{0}=-\frac{\hbar^{2}}{2m}\nabla^{2}-\mu$ is the
free Fermi gas Hamiltonian, with $m$ and $\mu$ as the electron mass
and chemical potential, respectively.

The 1-D wire is modeled as a Bosonized Luttinger liquid. Since the
literature on this subject is quite extensive (ref. [13] and
therein), we will only point out the more relevant details to the
topic at hand.

We represent the electronic fields\cite{Bos},
\begin{equation}
\psi_{r,\sigma}(x)=\frac{1}{\sqrt{2\pi\alpha}}e^{irk_{f}x}
e^{-i\sqrt{\frac{\pi}{2}}\left[r\phi_{c}(x)-\theta_{c}(x)+
\sigma\left(r\phi_{s}(x)-\theta_{s}(x)\right)\right]},
\end{equation}
in terms of boson charge and spin density fields:
$\partial_{x}\phi_{\nu}(x)$, and their conjugates,
$\theta_{\nu}(x)$, where $\nu=c/s$ denotes the charge and spin
sectors, respectively. In the above formula, $k_f$ is the Fermi wave
number, $\alpha$ is the short distance cutoff, $r$ is the chiral
index representing the right ($r=1$) and left ($r=-1$) moving part
of the electronic field, and $\sigma=1$ for $\uparrow$ spin while
$\sigma=-1$ for $\downarrow$ spin.

In terms of the boson fields, the action of the wire becomes a sum
of three contributions:
\begin{equation}
S_{W}=S_{c}(\theta_{c})+S_{s}(\phi_{s})
+S_{bs}(\phi_{s}),\label{Waction}
\end{equation}
where
\begin{widetext}
\begin{equation} S_{c}(\theta_{c})=\frac{1}{2}\int
dx\int_{0}^{\beta} d\tau
u_{c}K_{c}\left(\frac{1}{u_{c}^{2}}\left(\partial_{\tau}\theta_{c}
(x,\tau)\right)^{2}+\left(\partial_{x}\theta_{c}(x,\tau)\right)^{2}\right)
\end{equation}
\end{widetext}
describes the charge sector (after the dependence on the $\phi_c$
field has been integrated out\cite{Gi}). A similar expression
describes the spin sector, $S_s(\phi_s)$, with the subscript $c$
replaced by $s$ and $K_{c}$ replaced by $K_s^{-1}$. Here $K_{c}$,
$K_s$, $u_c$, and $u_s$ are model specific parameters describing the
interaction strength ($K$) and the mode velocity ($u$) of the charge
and the spin fields. For the noninteracting case $K_{c}=K_{s}=1$ and
$u_{c}=u_{s}=v_{f}$, where $v_{f}$ is the Fermi velocity of the
wire. For an interacting system, values of $K_{c}>1$ and $K_{s}<1$
correspond to attractive interactions, while $K_{c}<1$, $K_{s}>1$
correspond to repulsive ones. Generally in an interacting system the
velocity of the two modes differ, $u_{c}\neq u_{s}$.

The third term of the wire's action, $S_{bs}$, describes
backscattering of two electrons with opposite spins. Namely a
collision which effectively results in a spin flip between the right
and left moving parts of the electronic field
($\sim\psi^{\dagger}_{L,\sigma}\psi_{R,\sigma}\psi^{\dagger}_{R,-\sigma}\psi_{L,-\sigma}$).
This term has the form
\begin{equation}
S_{bs}=\frac{2g}{\left(2\pi\alpha\right)^{2}}\int
dxd\tau\cos\left(\sqrt{8\pi}\phi_{s}(x,\tau)\right),
\end{equation}
where $g$ is the backscattering coupling constant.

Finally, the tunneling between the superconductor and the wire is
described by
\begin{eqnarray}
S_{t}&=&\int d^4\xi' \int dx d\tau
\sum_{\sigma=\uparrow,\downarrow}
\bar{\varphi}_{\sigma}(\xi')t(\xi';x,\tau)
\psi_{\sigma}(x,\tau)\nonumber \\ & &{}+H.C.,
\end{eqnarray}
where $\psi_{\sigma}=\sum_r\psi_{r,\sigma}$, and $t(\xi';x,\tau)$ is
the tunneling matrix element. In the simplest case this tunneling is
instantaneous, homogeneous in space, short distant and SU(2) spin
symmetric. Under these assumptions it takes the form
\begin{equation}
t(\xi';x,\tau)=\tilde{t}(x)\delta(x'-x)\delta(y')\delta(z')\delta(\tau'-\tau)
\end{equation}
where $\tilde{t}(x)$ has the characteristic function form:
\begin{equation}
\tilde{t}(x)=\left\{ \begin{array}{l}
t_{0}\;\;\mbox{for} \;0\leq x\leq L\\
0\;\;\;\mbox{otherwise} \end{array}\right.\nonumber
\end{equation}
where $L$ is the length of the part of the wire embedded in the
superconductor (see fig.~\ref{pc_and_lc_junction} (b)).

The effective action of a wire embedded in a superconductor is
obtained by tracing out the superconductor degrees of freedom,
\begin{equation}
e^{-S_{eff}\left(\bar{\psi},\psi\right)} = \int
D[\bar{\varphi},\varphi]e^{-S\left(\bar{\varphi},
\varphi;\bar{\psi},\psi\right)}
\end{equation}
where $S$ is given by (\ref{action}). Since the action is quadratic
in the superconductor field, this integration is straightforward.
The result may be written as a sum of three terms,
$S_{eff}=S_{W}+S_{ph}+S_{pp}$, where $S_{ph}$, and $S_{pp}$
represent contributions  associated with Giever (particle-hole) and
Andreev (particle-particle) tunneling, respectively. At energies
above the superconductor gap $\Delta_{sc}$ the particle-hole term is
dominant and it's contribution, on integrating it out, will be to
renormalize the chemical potential. On the other hand, in the low
temperature regime (well below the superconductor gap $\Delta_{sc}$)
the particle-hole term vanishes and the main contribution comes from
the Andreev tunneling term, $S_{pp}$. In this limit, after averaging
over the rapid spatial oscillations, the tunneling becomes local in
space and time. Expressing it in terms of the bosonic fields we thus
have
\begin{eqnarray}
S_{eff}=S_{W}+S_{pp} \label{Seff}
\end{eqnarray}
where $S_{W}$ is given by (\ref{Waction}), and
\begin{equation}
S_{pp}\! =\! \frac{2\Delta}{\pi\alpha}\! \int \!dxd\tau \!
\cos \!\left(\!\sqrt{2\pi}\phi_{s}(x,\tau)\!\right)\! \cos
\left(\!\sqrt{2\pi}\theta_{c}(x,\tau)\!\right),\label{DVST}
\end{equation}
where $\Delta \simeq \left(
  t_{0}N_{0}\frac{\pi^{2}}{p_{f}}\right)^2$. Here  $N_{0}$
is the normal density of states of the SC at the Fermi level, and
$p_{f}$ is the Fermi momentum.

In terms of the original \textit{fermionic} fields, the tunneling term
has the form of the regular pairing term in the standard BCS theory:
\begin{eqnarray}
S_{pp}\!=\! \Delta\! \int \!dxd\tau \! \left(
\bar{\psi}_{\uparrow}(x,\tau)\bar{\psi}_{\downarrow}(x,\tau)
\!+\! \psi_{\downarrow}(x,\tau)\psi_{\uparrow}(x,\tau)\right)\nonumber
\end{eqnarray}

In the absence of electron-electron interactions, this term, along
with the free quadratic kinetic term of the model can be
diagonalized by the standard Bogolubov transformation. The
excitation spectrum of this system will be gapped with an energy of
$\Delta$.

\section{The Renormalization Group Flow Equations}

The effective action (\ref{Seff}) describes the physics of the
junction for a temperature up to the order of the superconductor
gap, $\Delta_{sc}$ (which will subsequently provide the high energy
cut-off of our system).

In order to describe the behavior of the system at much lower energy
scales, and take into account the electron-electron repulsive
interactions, we shall employ a real space RG approach, following
Giamarchi \& Schulz \cite{Gi,GS}. As usual in these cases the RG
procedure manifest itself in a flow of the coupling constants of the
problem as the ultraviolet cut off is reduced from $1/\alpha$ to
$1/\alpha'$. In our problem these coupling constants are: The
interaction strengths, $K_{\nu}$; The mode velocities, $u_{\nu}$;
The dimensionless backscattering constant, $y=\frac{g}{\pi u_{s}}$;
And the dimensionless pair tunneling strength
$\widetilde{\Delta}=\Delta \frac{\alpha}{u_{s}}$. The flow equations
of these coupling constants (see Appendix for the details of the
derivation) are:

\begin{eqnarray} \frac{dK_{c}}{dl} & = &
X_{c}\left(\frac{u_{s}}{u_{c}}\widetilde{\Delta}\right)^{2}
\label{DKCDL}\\
\frac{dK_{s}}{dl} & =
&-K_{s}^{2}\left(X_{s}\widetilde{\Delta}^{2}+\frac{y^{2}}{2}\right)
\label{DKSDL}\\
\frac{d\widetilde{\Delta}}{dl} & = &
\widetilde{\Delta}(2-\frac{1}{2}(K_{s}+K_{c}^{-1}+y))
\label{DDLDL}\\
\frac{dy}{dl}& = & y(2-2K_{s})-2X_{s}\widetilde{\Delta}^{2}
\label{DYDL}\\
\frac{du_{c}}{dl} & = &
u_{c}K_{c}^{-1}W_{c}\left(\frac{u_{s}}{u_{c}}\widetilde{\Delta}\right)^{2}
\label{DUCDL}\\
\frac{du_{s}}{dl} & = & u_{s}K_{s}W_{s}\widetilde{\Delta}^{2}
\label{DUSDL}
\end{eqnarray}
where $dl=d\log\alpha $ is the dimensionless change in the
ultraviolet cutoff,
\begin{widetext}
\begin{eqnarray}
X_{c (s)} & = & \frac{1}{2\pi}\int_{0}^{2\pi}d\varphi
\left(\cos^{2}(\varphi)+\left(\frac{u_{s (c)}}{u_{c
(s)}}\right)^{2}\sin^{2}(\varphi)\right)^{\frac{-g_{s (c)}}{2}}
\label{SFX}\\
W_{c (s)} & = & \frac{1}{2\pi}\int_{0}^{2\pi}d\varphi \cos(2\varphi)
 \times \left(\cos^{2}(\varphi)+\left(\frac{u_{s (c)}}{u_{c
(s)}}\right)^{2}\sin^{2}(\varphi)\right)^{\frac{-g_{s (c)}}{2}} \label{SFW}
\end{eqnarray}
and $g_{c}=K^{-1}_{c}$ while $g_{s}=K_{s}$.
\end{widetext}

We shall restrict our analysis to the spin symmetric case (i.e.
the situation where the interactions  between electrons
with parallel and opposite spins are identical). In this case\cite{GS}
 $K_{s}\simeq1+\frac{y}{2}$, and equations
(\ref{DKSDL}) and (\ref{DYDL}) for the dimensionless backscattering
$y$ and spin interaction strength $K_{s}$ reduce to:
\begin{equation}
\frac{dy}{dl}=-y^{2}-2X_{s}\widetilde{\Delta}^{2}, \label{DYDLSS}
\end{equation}
thus maintaining spin invariance.

\subsection{Analysis of the Flow Equations}
The above RG equations imply that the charge and spin velocities,
$u_c$ and $u_s$, renormalize towards each other. This is a
consequence of the correlations between spin and charge excitations
generated by the proximity effect. From Eqs. (\ref{DUCDL}),
(\ref{DUSDL}), and (\ref{SFW}) one can observe that the signs of
$W_{c}$ and $W_{s}$ are determined by the ratio $u_c/u_s$ in such a
way that the velocities approach each other.

Another consequence of the RG equations,(\ref{DKCDL}) and
(\ref{DKSDL}), is that $K_{c}$ can only grow, while $K_{s}$ can only
be reduced. This flow stem from the finite value of the tunneling
parameter $\widetilde{\Delta}$ (and $y$). Thus the RG behavior of
$\widetilde{\Delta}$ controls the behavior of the system.

\begin{figure}[h]
\begin{center}
\includegraphics[height=1.8in]{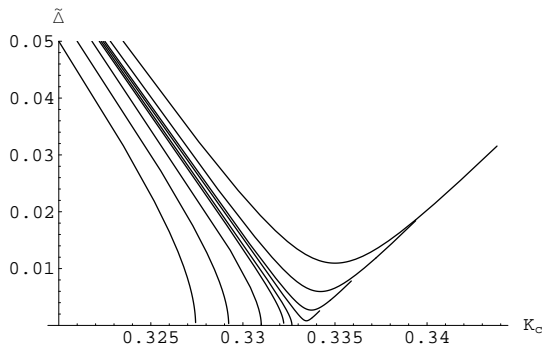}
\end{center}
\caption{Plots of the flow of $\widetilde{\Delta}$ as a function of
$K_{c}$ for initial values of $K^{(0)}_{c}$ between 0.32-0.325. The
transition between irrelevant ($\widetilde{\Delta}\rightarrow0$) and
relevant ($\widetilde{\Delta}\rightarrow\infty$) tunneling occurs at
$K^{(0)}_{c}\sim0.3223$. The initial values of
$\widetilde{\Delta}^{(0)}$ and $y^{(0)}$ are 0.05 and 0.1
respectively} \label{Kc_Delta_fig}
\end{figure}

\begin{figure}[h]
\begin{center}
\includegraphics[height=2.0in]{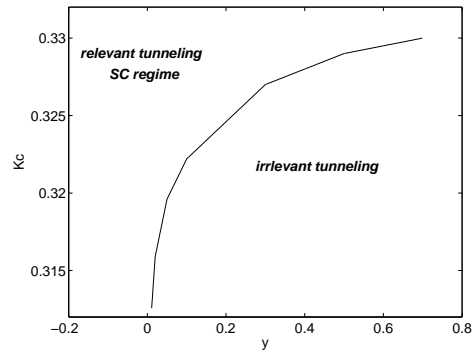}
\end{center}
\caption{The curve describes the separatix between relevant
tunneling, where the system flows to a superconducting state, and
irrelevant tunneling, where the system flows to a decoupled wire, as
a function of the bare values of $K_{c}$ and $y$}
\label{rel-irel_fig}
\end{figure}

The separatix between the regions where $\widetilde{\Delta}$ is
relevant or irrelevant can be obtained numerically for a given set
of bare parameters. It is clear that for initially repulsive
interactions ($K^{(0)}_{s}>1,\; y^{(0)}>0,\; K^{(0)}_{c}<1$),
$\widetilde{\Delta}$ will be relevant if:
\begin{equation}
4-(K^{(0)}_{s}+\left(K^{(0)}_{c}\right)^{-1}+y^{(0)})>0,
\end{equation}
since the flow equations can only drive $K_{c/s}$ to be more
"attractive". In Fig. \ref{Kc_Delta_fig} we depict the RG flow in
the $\widetilde{\Delta}$-$K_{c}$ plane, in the spin symmetric case,
for several initial values of $K_{c}^0$ in the range between
0.32-0.325. Fig.~\ref{rel-irel_fig} shows the separatix between
relevant and irrelevant tunneling for initial bare (repulsive)
values of $K_{c}$ and $y$. One can observe that $K_{c}$ must be
smaller than $\frac{1}{3}$ for the tunneling to be irrelevant.

In the case where $\widetilde{\Delta}$ is irrelevant, the wire
effectively decouples from the superconductor. The pair tunneling
between the two systems is suppressed and the superconductor-wire
junction becomes insulating. Notice however that this behavior takes
place at very strong repulsive interactions, $K_{c}<\frac{1}{3}$,
where the system tends to Wigner crystallize\cite{Gi}.

In the situation where $\widetilde{\Delta}$ is relevant, the system
flows to a singlet superconducting state. The interaction parameters
become attractive and spin-charge separation is no longer valid.
Yet, before the interaction parameters obtain their asymptotic
values (i.e $K_{c}\rightarrow\infty$, $K_{s}\rightarrow0$ and
$y\rightarrow-\infty$ which correspond to "infinitely" attractive
interactions), the RG equations  (\ref{DKCDL}- \ref{DUSDL}) will
cease to be valid since the small parameters of our perturbation
theory, $\widetilde{\Delta}$ and $y$, will flow to the strong
coupling regime.

\subsection{Length Dependance of the Effective Gap}

In what follows we consider a weakly interacting Luttinger liquid,
where the bare interaction parameters are close to unity. Moreover,
in order to ensure that the perturbative RG equations (\ref{DKCDL}-
\ref{DUSDL}) remain valid we shall assume that $L$, the embedding
length,  is of order or smaller then the bare superconducting
correlation length, $v_f/\Delta$. Since $L$ serves as an infrared
cutoff for the RG flow, this condition ensures that the this flow is
confined to the perturbative regime. We shall also assume that the
temperature is much smaller than $\Delta$, and approximate the
velocities of the spin and charge sectors by their asymptotic
renormalized values: $u_{c}=u_{s}=v_{f}$.

In the limit of weak interactions the RG equations for $K_{s/c}$
are of second order in the perturbation parameters ($\tilde{\Delta}
\ll  1, y\ll 1$), while the
equation for $\widetilde{\Delta}$ (Eq. \ref{DDLDL}) is of first order
(neglecting the $y$ dependance, which is of second order as well).
The equation for the coupling $y$ (Eq.~\ref{DYDL}) is also second
order in small parameters, at least if we consider the spin symmetric case .
Thus one can assume $K_c$, $K_s$, and $y$ to be
approximately constants, and consider the simplified equation for
$\widetilde{\Delta}$:
\begin{eqnarray}
\frac{d\widetilde{\Delta}}{dl} & = &
\gamma \widetilde{\Delta}, \label{RGdelta}
\end{eqnarray}
where
\begin{eqnarray}
\gamma = 2-\frac{1}{2}(K_{s}+K_{c}^{-1}). \label{beta}
\end{eqnarray}

Integrating the above equations from $l=0$ to $l=\log(L/\alpha)$, and
using the relation $\tilde{\Delta}= \Delta \alpha/v_f$, we
obtain the renormalized value of the gap of the wire embedded
in the superconductor:
\begin{eqnarray}
\Delta_{eff}= \Delta \left(\frac{L}{\alpha}\right)^{\gamma-1}.
\end{eqnarray}
In particular, repulsive interactions ($K_c<1$ and $K_s>1$, and therefore
$\gamma-1<0$) reduce the effective gap in the wire.

This renormalization of the gap implies that the effective correlation
length,
\begin{eqnarray}
\xi_{eff}=\frac{v_f}{\Delta_{eff}}= \xi \left(\frac{L}{\alpha}
\right)^{1-\gamma} \label{eff-cor}
\end{eqnarray}
 is larger than the bare correlation
length $\xi = v_f/\Delta$.

The above results are valid as long as the RG flow stays within the
perturbative regime, namely $\tilde{\Delta}<1$. This condition
implies that $L$ should be shorter than $\xi (\xi/\alpha)^{\frac{1}{\gamma}-1}$.

\section{The Wire-Superconductor Junction}

At this stage of the RG procedure (pursued up to the scale $L$) the
junction between the wire and the superconductor may be assumed to
be point-like. Thus if the temperature (or the applied voltage) is
smaller than $\hbar v_f/L$ one may continue to integrate out the
high energy degrees of freedom in the part of the wire which is not
embedded into the superconductor down to the relevant energy scale.
This may be achieved following the procedure described in the
literature\cite{TK2}. The important ingredient, now, is the
magnitude of normal back scattering from the junction. The latter is
of order of $r_{N}\simeq r_0 e^{-2 L/\xi_{eff}}$ where $r_0$ is
system specific reflection amplitude in the absence of
superconductivity, while $\xi_{eff}$ is the correlation length
(\ref{eff-cor}) within the part of the wire embedded in the
superconductor. This behavior of $r_N$ results form the fact that
only the charge that is not converted to the condensate backscatters
from the edge of the wire embedded in the superconductor. According
to\cite{BTK} the normal current reduces exponentially with the
distance on a length scale of $\xi$ , which gives the above estimate
for $r_N$. Now, since $\xi_{eff}$ depends on $L$, the magnitude of
the normal reflection has a stretched exponential dependence:
\begin{eqnarray}
r_{N}\simeq r_0 e^{-a L^\gamma}
\end{eqnarray}
where $a=2 \alpha^{1-\gamma}/\xi$, and the amount of stretch, $\gamma$,
is dictated by the nature of the electron-electron interactions
within the wire, as follows from (\ref{beta}).

The behavior of the backscattering, clearly, manifests itself in the
conductance of the junction. In the limit of sufficiently high
temperatures, one obtains \cite{TK1}:
\begin{eqnarray}
G & = & G^{0}_{NS}-\delta G, \label{DSG0MINDG}
\end{eqnarray}
where $G^{0}_{NS}=2G^{0}_{NN}=4\frac{e^2}{2\pi\hbar}$ represents the
conductance of an ideal junction where only perfect Andreev reflections take
place, and
\begin{eqnarray}
\delta G & \propto & r_{0}^2e^{-2a L^\gamma} ~T^{-2(1-K_{c})}. \label{DSSPCR}
\end{eqnarray}

The above  formula holds for the range of temperature where $\delta
G \ll G_{NS}^0$,  since $\delta G$ cannot be larger than $G^{0}_{NS}$.
Nevertheless, it has been shown \cite{FHO}-\cite{ACZ}, that any
scatterer, at a point contact between a wire with repulsive
interactions and a SC, will eventually drive the conductance to zero
as the temperature is lowered. Since any finite length junction will
have some backscattering, the conductance should drop to zero for
low enough temperatures, as illustrated in Fig.~4. The crossover
temperature, $T^*$ from the conductive and the insulating behavior
of the junction depends on the length of the wire embedded in the
superconductor and its scaling behavior may be deduced from
(\ref{DSG0MINDG}) and (\ref{DSSPCR}):
\begin{equation}
T^{*}\propto e^{-\frac{a}{1-K_{c}}L^{\gamma}} \label{DCTSTR}
\end{equation}

This result implies that the temperature scale at which the effects of
backscattering becomes substantial reduce as a stretched exponent
with the length of the junction, and the stretch is determined by
the interactions, through the parameter $\gamma$.

\begin{figure}[ht!]
\begin{center}
\includegraphics[height=2in , width=2.5in]{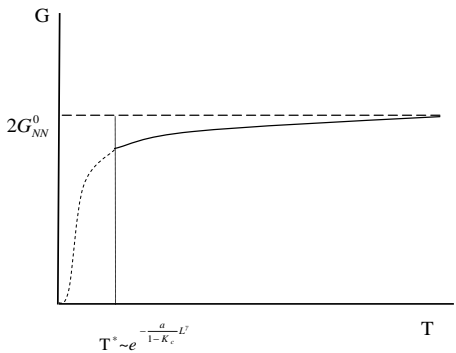}
\end{center}
\caption{A schematic graph of the conductance as a function of
temperature for repulsive interactions in the wire. The decay of the
conductance depends on the length of the junction as a stretch
exponent. The longer the junction, the lower the temperature at
which the decay sets in. In the limit of weak backscattering,
represented by the solid curve, we can estimate the temperature
dependance of the conductance through Eq. (\ref{DSSPCR}). In the
zero temperature limit, represented by the dashed curve, any initial
finite backscattering will eventually drive the conductance to zero.
The crossover temperature, Eq. (\ref{DCTSTR}) between the two
behaviors is estimated to scale as a stretched exponent in the
length of the junction.} \label{schem_fig}
\end{figure}.

\section{Summary and Conclusions}

In this work we studied a junction of a 1-D wire embedded a certain
length, $L$, into a bulk superconductor. We first characterized the
nature of the contact between the 1-D wire and the superconductor
using a real space RG scheme. We found that repulsive interactions
in the wire compete against the superconducting order being imposed
by the bulk superconductor. The system can  flow to either of two
phases, depending on the nature of the interactions. When the
interactions are
strongly repulsive the tunneling between the two systems becomes
irrelevant and the wire essentially decouples from the bulk
superconductor. For moderate repulsive interactions, and for
attractive interactions, tunneling is  relevant, and the bulk
superconductor induces superconducting order in the wire. The gap
opened in the wire depend on the tunneling strength, and
electron-electron interactions modify its nominal value.

The finite length of the part of the wire embedded in the
superconductor, $L$, implies that the RG flow, in general, does not
reach its asymptotic (non-perturbative) limit. Thus $L$ introduces
itself in the behavior of the effective gap, and the effective
correlation length, in the wire. This, in turn, dictates a stretched
exponential behavior of the normal reflection from the junction as
function of $L$.  For energy scales (temperature or voltages)
beneath the effective gap, we described the qualitative picture of
the conductance as a function of temperature and length of the
embedded segment of wire, see Fig. \ref{schem_fig}.

Several simplifications have been used for our analysis: One is that
our model treats a semi-infinite wire with a single junction, while
in practical situations a finite wire is usually connected to two
reservoirs.  This idealization holds as long as the segment of the
wire outside the superconductor is long enough compared to $\hbar
v_f/T$, where $v_f$ is the Fermi velocity and $T$ is the
temperature.  Additional simplification is the assumption that the
embedding of the wire into the bulk superconductor does not
introduce inhomogeneities, i.e. the wire can be still considered to
be clean, and that the tunneling to the superconductor is homogenous
along the wire. This approximation holds when  the transport mean
free path in the wire is longer than $L$. In the opposite limit one
expects a different behavior of the proximity effect which will
change the conductance of the system.

The authors gratefully acknowledge discussions with Dror Orgad. This
work has been supported in part by the Israel Science Foundation
(ISF) funded by the Israeli Academy of Science and Humanities, and
by the USA-Israel Binational Science Foundation (BSF).

\section{Appendix}

The mathematical formulation used in this work follows closely the
real space RG procedure used by Giamarchi \& Schulz \cite{Gi,GS}: In
this procedure one evaluates a correlation function in the wire, of
the form (the time ordering symbol is suppressed):
\begin{equation}
R_{\varphi}(x_{a},\tau_{a};x_{b},\tau_{b})=\left\langle
e^{i\gamma\sqrt{2\pi}\left(\varphi(x_{a},\tau_{a})-\varphi(x_{b},\tau_{b})\right)}\right\rangle,
\end{equation}
where $\varphi$ can symbolize any of the boson fields and $\gamma$
is some constant. For the following discussion it will suffice to
examine only one of the sectors, for instance the spin sector. The
same considerations can be carried on straightforwardly for the
charge sector.

The fact that the relevant boson fields in the spin sector is
$\phi_{s}$ leads naturally to the evaluation of the correlation
functions:
\begin{equation}
R_{\phi_{s}}(x_{a},\tau_{a};x_{b},\tau_{b})= \left\langle
e^{i\gamma\sqrt{2\pi}\left(\phi_{s}(x_{a},\tau_{a})-\phi_{s}(x_{b},\tau_{b})\right)}\right\rangle.
\end{equation}
Unfortunately this correlation function cannot be calculated exactly
using the complete effective action (\ref{Seff}). Though, if the tunneling
and backscattering parameters, ($t$ and $g$ respectively), are small
then it may be computed perturbatively. To second order in these
parameters this function is found to be:
\begin{equation}
R_{s}(\vec{r}_{a,b})=
e^{-\gamma^{2}\left(K^{eff}_{s}F_{s}(\vec{r}_{a,b})+D^{eff}_{s}\sin^{2}(\varphi_{\vec{r}_{a,b},s})\right)},
\label{DVREF}
\end{equation}
where $\vec{r}_{a,b}=\vec{r}_{a}-\vec{r}_{b}$, and $\varphi_{\vec{r},s}$
is the angle between the vector $\vec{r}=(x,u_{s}\tau)$ and the $x$
axis. The function $F_{s}$ is (at zero temperature) \cite{Gi}:
\begin{equation}
F_{s}(x,\tau)=\frac{1}{2}\ln\left(\frac{x^{2}+\left(u_{s}|\tau|+\alpha\right)^{2}}{\alpha^{2}}\right).
\label{DVFNU}
\end{equation}
Apart from the term proportional to
$\sin^{2}(\varphi_{\vec{r}_{a,b},s})$, this functional form is
identical to the free correlation function:
\begin{equation}
R_{s}^{(0)}(\vec{r}_{a}-\vec{r}_{b})=
e^{-\gamma^{2}K_{s}F_{s}(\vec{r}_{a}-\vec{r}_{b})},
\end{equation}
but with an effective Luttinger interaction constant $K_{s}^{eff}$
modified by the perturbations:
\begin{widetext}
\begin{eqnarray}
K_{s}^{eff}&=&K_{s}-K_{s}^{2}\left(\widetilde{\Delta}^{2}X_{s}
\int_{\alpha}^{\infty}\frac{dr}{\alpha}\left(\frac{r}{\alpha}\right)^{3-K_{s}-K_{c}^{-1}}+
\frac{y^{2}}{2}\int\frac{dr}{\alpha}\left(\frac{r}{\alpha}\right)^{3-4K_{s}}\right).
\end{eqnarray}
\end{widetext}
Here $\widetilde{\Delta}=\frac{\Delta\alpha}{u_{s}}$ is the
dimensionless tunneling parameter, $y=\frac{g}{\pi u_{s}}$ is the
dimensionless spin backscattering parameter, and $X_{s}$ is a
geometrical term given by:
\begin{equation}
X_{s} \!=\! \frac{1}{2\pi}\!\int_{0}^{2\pi}\!\!d\varphi
\left(\cos^{2}(\varphi)\!+\!\left(\frac{u_{c}}{u_{s}}\right)^{2}\sin^{2}(\varphi)\right)^{\frac{-K^{-1}_{c}}{2}}
\end{equation}

Integrating out high energy degrees of freedom, near the ultraviolet cutoff,
corresponds to integrating out a "small ring" between
$\alpha\rightarrow\alpha'=\alpha+d\alpha$, where $\alpha$ is the
small distances parameter of the model. After integration and
rescaling, an infinitesimal change is generated in the
expressions for $K_{s}^{eff}$. In order to keep
$K_{s}^{eff}$ constant  with the reduction of the cutoff,  it is
required that the bare parameters change. For instance, we find that:
\begin{equation}
K_{s}(\alpha')  \!=\!  K_{s}(\alpha)\!-\!
K_{s}^{2}(\alpha)\left(\!
  \widetilde{\Delta}^{2}(\alpha)X_{s}(\alpha)\!+\!\frac{y^{2}(\alpha)}{2}\! \right)\frac{d\alpha}{\alpha}
\label{DVKS}
\end{equation}
which generates an (exact) differential equation for $K_{s}$:
\begin{equation}
\frac{dK_{s}}{dl}=-K_{s}^{2}(l)\left(X_{s}\widetilde{\Delta}^{2}(l)+\frac{y^{2}(l)}{2}\right)
\end{equation}
(here $dl=\frac{d\alpha}{\alpha}$).

In a similar fashion one obtains differential equations for the
parameters $\widetilde{\Delta}$ and $y$:
\begin{eqnarray}
\frac{d\widetilde{\Delta}^{2}}{dl} & = &
\widetilde{\Delta}^{2}(4-K_{s}-K_{c}^{-1}),
\label{DVRT}\\
\frac{dy^{2}}{dl} & = & y^{2}(4-4K_{s}).
\label{DVRY}
\end{eqnarray}

The $\sin^{2}(\varphi_{\vec{r}_{a,b},s})$ contribution to the
correlation function arises from the fact that the tunneling
perturbation couples the spin and charge sectors, which were
uncoupled without this term. Mathematically, this term characterizes
the anisotropy between the space ($x$) and time ($u_{s}\tau$)
directions. It's pre-factor $D^{eff}$ is given by:
\begin{equation}
D_{s}^{eff}=D_{s}+K_{s}^{2}\widetilde{\Delta}^{2}W_{s}
\int_{\alpha}^{\infty}\frac{dr}{\alpha}\left(\frac{r}{\alpha}\right)^{3-K_{s}-K_{c}^{-1}},
\end{equation}
where $W_{s}$ is another geometric term factor:
\begin{eqnarray}
W_{s}\!=\! \frac{1}{2\pi}\!\int_{0}^{2\pi}\!\!\! d\varphi \cos(2\varphi)
\left(\!\cos^{2}(\varphi)\!+\!\left(\frac{u_{c}}{u_{s}}\right)^{2}\!\!
  \sin^{2}(\varphi)\!\right)^{\frac{-K^{-1}_{
c}}{2}}. \nonumber
\end{eqnarray}

Applying the same renormalization scheme to $D^{eff}$, will generate
the flow of the parameter $D_{s}$:
\begin{equation}
\frac{dD_{s}}{dl}=K_{s}^{2}(l)W_{s}(l)\widetilde{\Delta}^{2}(l).
\end{equation}
It should be noted that $D_{s}$ is initially zero but is generated
under renormalization.

The parameter $D_{s}$ controls the renormalization of the velocity
parameter. As long as the space and time directions are isotropic
the velocity parameter does not flow under renormalization, but they
should flow in the anisotropic case. Indeed, assuming that initially
the correlation function is described by the function
$K_{s}F_{s}(\vec{r}_{a,b})$, then a small change of $du_{s}$ will
generate the term:
$\;\frac{K_{s}}{u_{s}}\sin^{2}(\varphi_{\vec{r}_{a,b},s})\cdot
du_{s}\;$. Up to a factor, this is exactly the anisotropy term.
Therefore, the renormalization of $D_{s}$ is equivalent to that of
the velocity $u_{s}$ by the following relation (e.g. Eq.
\ref{DVREF}):$\frac{du_{s}}{dl}=\frac{u_{s}}{K_{s}}\frac{dD_{s}}{dl}$.

The flow equations for the charge sector can be obtained by an
identical procedure. The equations obtained are:
\begin{eqnarray}
\frac{dK_{c}}{dl} & = &
=X_{c}(l)\left(\frac{u_{s}}{u_{c}}\widetilde{\Delta}(l)\right)^{2},\\
\frac{dD_{c}}{dl} & = &
K_{c}^{-2}(l)W_{c}(l)\left(\frac{u_{s}}{u_{c}}\widetilde{\Delta}(l)\right)^{2}.
\end{eqnarray}

Finally these equations cannot be exactly correct, since they do not
maintain the spin invariance SU(2) of a model that was spin
invariant to begin with. (A spin symmetric model is one where the
interactions between electrons of opposite and parallel spin are
identical). Since the perturbations do not break this symmetry,
something in the above result is insufficient. Indeed, it turns out
that the remedy for this problem lies in the inclusion of the third
order terms of perturbation theory. This correction is presented in
reference \cite{GS}. It affects only the equations for
$\widetilde{\Delta}$ (Eq. (\ref{DVRT}))and $y$ (Eq. (\ref{DVRY}))
which become:
\begin{eqnarray}
\frac{d\widetilde{\Delta}}{dl} & = &
\widetilde{\Delta}(2-\frac{1}{2}(K_{s}+K_{c}^{-1}+y)), \\
\frac{dy}{dl}& = & y(2-2K_{s})-2X_{s}\widetilde{\Delta}^{2}.
\end{eqnarray}


\begin{thebibliography}{99}

\bibitem{KF}
C. L. Kane and M. P. A. Fisher, Phys. Rev B {\bf 46}, 15233 (1992).

\bibitem{FN}
A. Furusaki and N. Nagaosa, Phys. Rev B {\bf 47}, 4631 (1993).

\bibitem{YGM}
D. Yue, L. I. Glazman and K. A. Matveev, Phys. Rev. B {\bf 49}, 1966 (1994).

\bibitem{Cndct} I. Safi and H.J. Schulz, Phys. Rev. B {\bf 52}, R17040
  (1995); D.L. Maslov and M. Stone, Phys. Rev. B {\bf 52}, R5539
  (1995); Y. Oreg and A. M. Finkel'stien, cond-mat/9607149 v1 (1996).

\bibitem{BTK}
G. E. Blonder, M. Tinkham and T.M. Klapwijk, Phys. Rev. B {\bf 25},
4515 (1982)

\bibitem{FHO}
R. Fazio, F. W. J. Hekking and A. A. Odintsov, Phys. Rev. Lett. {\bf
74}, 1843 (1995).

\bibitem{MSG}
D. L. Maslov, M. Stone, P. M. Goldbart and D. Loss, Phys.
Rev. B {\bf 53} 1548 (1996)

\bibitem{TK1}
Y. Takane and Y. Koyama, J. Phys. Soc. Japan {\bf 65}, 3630 (1996).

\bibitem{TK2}
Y. Takane and Y. Koyama, J. Phys. Soc. Japan {\bf 66}, 419 (1997).

\bibitem{ACZ}
I. Affleck, J. S. Caux and A. M. Zagoskin, Phys. Rev. B 62
{\bf 1433} (2000).

\bibitem{VBBF}
S. Vishveshwara, C. Bena, L. Balents and M. P. A. Fisher,
Phys. Rev. B {\bf 66}, 165411 (2002).

\bibitem{Bos}
There are actually quite a few different conventions used in the
bosonization literature. Apart from a factor of $\sqrt{\pi}$ in the
normalization of the bosonic fields, we follow the one used by
\cite{Gi}.

\bibitem{Gi}
T. Giamarchi, \textit{Quantum Physics in One Dimension}. Clarendon
Press, Oxford (2004).

\bibitem{GS}
T. Giamarchi and H. J. Schulz, Phys. Rev. B {\bf 37}, 325 (1988).



\end{thebibliography}
\end{document}